\documentclass[journal,twoside]{IEEEtran}
\usepackage[T1]{fontenc}
\usepackage{cite}
\usepackage{amsmath,amssymb,amsfonts}
\usepackage{algorithmic}
\usepackage[]{graphicx}
\usepackage[mathlines,switch]{lineno}
\usepackage{multicol,multirow}
\usepackage{epsfig}
\usepackage{TUSON}
\usepackage{balance}

\DeclareUnicodeCharacter{00A0}{\nobreakspace}
\DeclareUnicodeCharacter{2008}{\,}
\DeclareUnicodeCharacter{2009}{\,}
\DeclareUnicodeCharacter{200A}{\,}
\DeclareUnicodeCharacter{2010}{-}
\DeclareUnicodeCharacter{2013}{\textendash}
\DeclareUnicodeCharacter{2014}{\textemdash}
\DeclareUnicodeCharacter{2212}{\textminus}

\DeclareUnicodeCharacter{2019}{\textquoteright}
\DeclareUnicodeCharacter{201C}{\textquotedblleft}
\DeclareUnicodeCharacter{201D}{\textquotedblright}
\DeclareUnicodeCharacter{2022}{\textbullet}
\DeclareUnicodeCharacter{00B7}{\textperiodcentered}
\DeclareUnicodeCharacter{2215}{/}
\DeclareUnicodeCharacter{FB01}{fi}

\DeclareUnicodeCharacter{00A9}{\textcopyright}
\DeclareUnicodeCharacter{00AE}{\textregistered}
\DeclareUnicodeCharacter{00B0}{\textdegree}
\DeclareUnicodeCharacter{00B1}{\textpm}
\DeclareUnicodeCharacter{00B2}{\ensuremath{^{2}}}
\DeclareUnicodeCharacter{00B5}{\textmu}
\DeclareUnicodeCharacter{00BA}{\textordmasculine}
\DeclareUnicodeCharacter{00D7}{\texttimes}
\DeclareUnicodeCharacter{2080}{\ensuremath{_{0}}}
\DeclareUnicodeCharacter{2082}{\ensuremath{_{2}}}
\DeclareUnicodeCharacter{2083}{\ensuremath{_{3}}}
\DeclareUnicodeCharacter{209C}{\ensuremath{_{\mathrm{t}}}}
\DeclareUnicodeCharacter{2223}{\ensuremath{\mid}}
\DeclareUnicodeCharacter{2248}{\ensuremath{\approx}}
\DeclareUnicodeCharacter{223C}{\ensuremath{\sim}}
\DeclareUnicodeCharacter{2264}{\ensuremath{\leq}}
\DeclareUnicodeCharacter{25CB}{\ensuremath{\circ}}

\DeclareUnicodeCharacter{03A9}{\ensuremath{\Omega}}
\DeclareUnicodeCharacter{03B1}{\ensuremath{\alpha}}
\DeclareUnicodeCharacter{03B5}{\ensuremath{\varepsilon}}
\DeclareUnicodeCharacter{03BB}{\ensuremath{\lambda}}
\DeclareUnicodeCharacter{03BC}{\textmu}
\DeclareUnicodeCharacter{03BD}{\ensuremath{\nu}}
\DeclareUnicodeCharacter{03C9}{\ensuremath{\omega}}
\DeclareUnicodeCharacter{03F5}{\ensuremath{\epsilon}}

\DeclareUnicodeCharacter{1D452}{\ensuremath{e}}
\DeclareUnicodeCharacter{1D458}{\ensuremath{k}}
\DeclareUnicodeCharacter{1D680}{\ensuremath{\mathtt{Q}}}

\DeclareUnicodeCharacter{00C1}{\'{A}}
\DeclareUnicodeCharacter{00E1}{\'{a}}
\DeclareUnicodeCharacter{00E3}{\~{a}}
\DeclareUnicodeCharacter{00E4}{\"{a}}
\DeclareUnicodeCharacter{00E5}{\aa}
\DeclareUnicodeCharacter{00E7}{\c{c}}
\DeclareUnicodeCharacter{00E9}{\'{e}}
\DeclareUnicodeCharacter{00EB}{\"{e}}
\DeclareUnicodeCharacter{00EF}{\"{\i}}
\DeclareUnicodeCharacter{00F3}{\'{o}}
\DeclareUnicodeCharacter{00F6}{\"{o}}
\DeclareUnicodeCharacter{00F8}{\o}
\DeclareUnicodeCharacter{00FC}{\"{u}}
\DeclareUnicodeCharacter{00FF}{\"{y}}
\DeclareUnicodeCharacter{0117}{\.{e}}
\DeclareUnicodeCharacter{0142}{\l}
\DeclareUnicodeCharacter{017D}{\v{Z}}

\newcommand{\tw}[1]{\textbf{#1}}

\title{Thickness-Field-Excited Bimorph Piezoelectric Micromachined Ultrasonic Transducer in Submicrometer Thin-Film Lithium Niobate}

\author{Vakhtang~Chulukhadze,~ \IEEEmembership{Student~Member,~IEEE,} Jiajun~Huang, Zihuan~Liu,~\IEEEmembership{Student~Member,~IEEE,} Will~Leskowitz, Tzu-Hsuan~Hsu,~\IEEEmembership{Member,~IEEE,} Yuchen Ma, ~\IEEEmembership{Student~Member,~IEEE,} Udit~Rawat,~\IEEEmembership{Member,~IEEE,} Yao~Yu, Bichoy~Bahr,~\IEEEmembership{Member,~IEEE,} Neal~A.~Hall, and~Ruochen~Lu,~\IEEEmembership{Senior~Member,~IEEE}%
\thanks{Manuscript received [date], revised [date], accepted [date].  This work was supported in part by the Defense Advanced Research Projects Agency (DARPA) High Operational Temperature Sensors (HOTS) project under Award HR00112420334 and in part by the National Institute on Deafness and Other Communication Disorders (NIDCD) of the National Institutes of Health (NIH) under Award R41DC022811.}%
\thanks{V. Chulukhadze, J. Huang, Z. Liu, W. Leskowitz, T.-H. Hsu, Y. Ma, N. A. Hall, and R. Lu are with the Chandra Family Department of Electrical and Computer Engineering, The University of Texas at Austin, Austin, TX 78712 USA (e-mail: vatoc@utexas.edu).}%
\thanks{U. Rawat, Y. Yu, and B. Bahr are with Kilby Labs, Texas Instruments, Dallas, TX 75243 USA.}}

\IEEEaftertitletext{\GA{figures/TUFFC_TFE_GA}

\Abstract{\begin{abstract}

This work presents the first demonstration of a submicrometer-thick thickness-field-excited (TFE) lithium niobate (LN) flexural-mode piezoelectric micromachined ultrasonic transducer (PMUT). A 670-nm-thick bilayer periodically poled piezoelectric film (P3F) 128Y LN is used to demonstrate a suspended membrane with an area of 0.25~mm$^{2}$, achieving a low resonance frequency of 33.4~kHz for airborne transduction. Its electrical measurements reveal an extracted effective electromechanical coupling efficiency (\(k_{\mathrm{eff}}^{2}\)) of 3\% and a moderate static capacitance \(C_0\) of 17~pF. Meanwhile, mechanical measurements using a laser Doppler vibrometer (LDV) yield a high quasi-static peak out-of-plane displacement (\(\xi_{\mathrm{static}}\)) of 145~nm/V and a high corresponding open-circuit sensitivity (\(S_{\mathrm{OC}}\)) of 1.4~mV/Pa in the audio band. The fabricated device is compared with state-of-the-art (SoA) airborne piezoelectric transducers with sub-100-kHz resonance frequencies, underscoring the potential of the TFE P3F LN transducer platform by simultaneously demonstrating strong actuation capability while featuring high sensitivity. 
\hblne
\end{abstract}}
\vspace{1\baselineskip}\vspace*{-1pt}
}

\begin{document}

\bstctlcite{IEEEexample:BSTcontrol}

\maketitle

\Highlights[1]{This work presents the first demonstration of a thickness-field-excited (TFE) lithium niobate (LN) clamped-plate flexural-mode PMUT with a submicrometer-thick active layer.}

\Highlights[2]{Measurements reveal a high extracted effective electromechanical coupling efficiency ($k_{\mathrm{eff}}^{2}$) of 3\%, alongside high actuation and sensitivity metrics in the audio band, such as a \(\xi_{\mathrm{static}}\) of 145~nm/V and flat-band \(S_{\mathrm{OC}}\) of 1.4~mV/Pa.}

\Highlights[3]{The resulting TFE transducer exhibits high performance and integrability with a compact form factor, indicating promise as an airborne piezoelectric transducer platform.}


\PrintHighlights

\section{Introduction}
\label{sec:introduction}

\IEEEPARstart{P}{iezoelectric} micromachined ultrasonic transducers (PMUTs) are frequently employed as passive components in airborne applications that require a combination of high performance, compact footprint, and low-power operation, such as range-finding and microphones~\cite{luo_airborne_2021, peng_9-meter-long_2024, xia_high-spl_2023, horowitz_development_2007, williams_aln_2012}. Incumbent piezoelectric material platforms, e.g., lead zirconate titanate (PZT), potassium sodium niobate (KNN), and scandium aluminum nitride (ScAlN), offer distinct trade-offs in actuation and sensing, but remain limited in bidirectional performance (Table~\ref{tab:fom_comparison})~\cite{belgacem_piezoelectric_2006, caro_piezoelectric_2015, shi_enhanced_2020}. 

In an effort to accurately evaluate material-level performance for flexural-mode operation, the effective electromechanical tensor components under the stress-charge piezoelectric constitutive equation system can be utilized. These parameters reflect the mixed boundary conditions of the PMUT, being clamped in-plane while free to deform out-of-plane~\cite{dubois_measurement_1999, mertin_enhanced_2017}. The resulting effective stress-charge piezoelectric constant (\(e_{ij,\mathrm{eff}}\), in C/m\(^2\), where \(i\) is the electric field direction and \(j\) is the stress component) acts as a bidirectional figure of merit (FoM) when combined with the effective electrical permittivity (\(\varepsilon_{ii,\mathrm{eff}}\)) and the loss tangent (\(\tan\delta\))~\cite{muralt_which_2017,lu_piezoelectric_2020}. Accordingly, the comparison in Table~\ref{tab:fom_comparison} exhibits concurrent limitations in incumbent piezoelectric material platforms: PZT features a large \(e_{31,\mathrm{eff}}\), exhibiting potential for efficient actuation, but remains limited as a sensor due to its high \(\varepsilon_{ii,\mathrm{eff}}\) and \(\tan\delta\). On the other hand, ScAlN is well-positioned as a sensor, but features limited actuation capability, highlighting the need for an alternative PMUT platform with high bidirectional performance metrics. 

\begin{table}[!t]
\caption{Piezoelectric transducer FoM comparison between incumbent piezoelectric platforms, highlighting the high FoM paired with a moderate \(C_0\) density in 128Y LN.}
\label{tab:fom_comparison}
\centering
\renewcommand{\arraystretch}{1.2}
\setlength{\tabcolsep}{2pt}
\footnotesize
\resizebox{\columnwidth}{!}{%
\begin{tabular}{lccccc}
\hline
Material & \begin{tabular}{@{}c@{}}Exci-\\tation\end{tabular} & \begin{tabular}{@{}c@{}}$e_{ij,\mathrm{eff}}$\\(C/m$^{2}$)\end{tabular} & \begin{tabular}{@{}c@{}}$e_{ij,\mathrm{eff}}/(\varepsilon_{ii,\mathrm{eff}}\tan\delta)^{1/2}$\\($10^{6}$(J/m$^{3}$)$^{1/2}$)\end{tabular} & \begin{tabular}{@{}c@{}}Capacitance\\Density\end{tabular} & \begin{tabular}{@{}c@{}}Interface\\Electrode\end{tabular} \\
\hline
Epitaxial PZT & TFE & 24.0 & 5.2 & Very High & Y \\
Sputtered KNN & TFE & 10.0 & 1.7 & Very High & Y \\
Sputtered AlN & TFE & 1.0 & 3.0 & Moderate & Y \\
Sputtered Sc\(_{36}\)Al\(_{64}\)N & TFE & 2.4 & 3.8 & Moderate & Y \\
X-cut LN & LFE & 5.3 & 13.4 & Low & N \\
\tw{128Y LN} & \tw{TFE} & \tw{3.6} & \tw{9.0} & \tw{Moderate} & \tw{N} \\
\hline
\end{tabular}}
\par\smallskip
\parbox{\columnwidth}{\footnotesize The subscripts are \(31\) for TFE (\(3,1\)) excitation and \(11\) for LFE (\(1,1\)) excitation. Magnitudes are listed; the sign of each coefficient depends on the material and orientation. Values from~\cite{qiao_full_2020, shi_enhanced_2020, caro_erratum_2015, wolff_al1xscxn_2022, streque_stoichiometric_2019}.}
\end{table}

\begin{figure}[!t]
    \centering
    \includegraphics[width=\columnwidth]{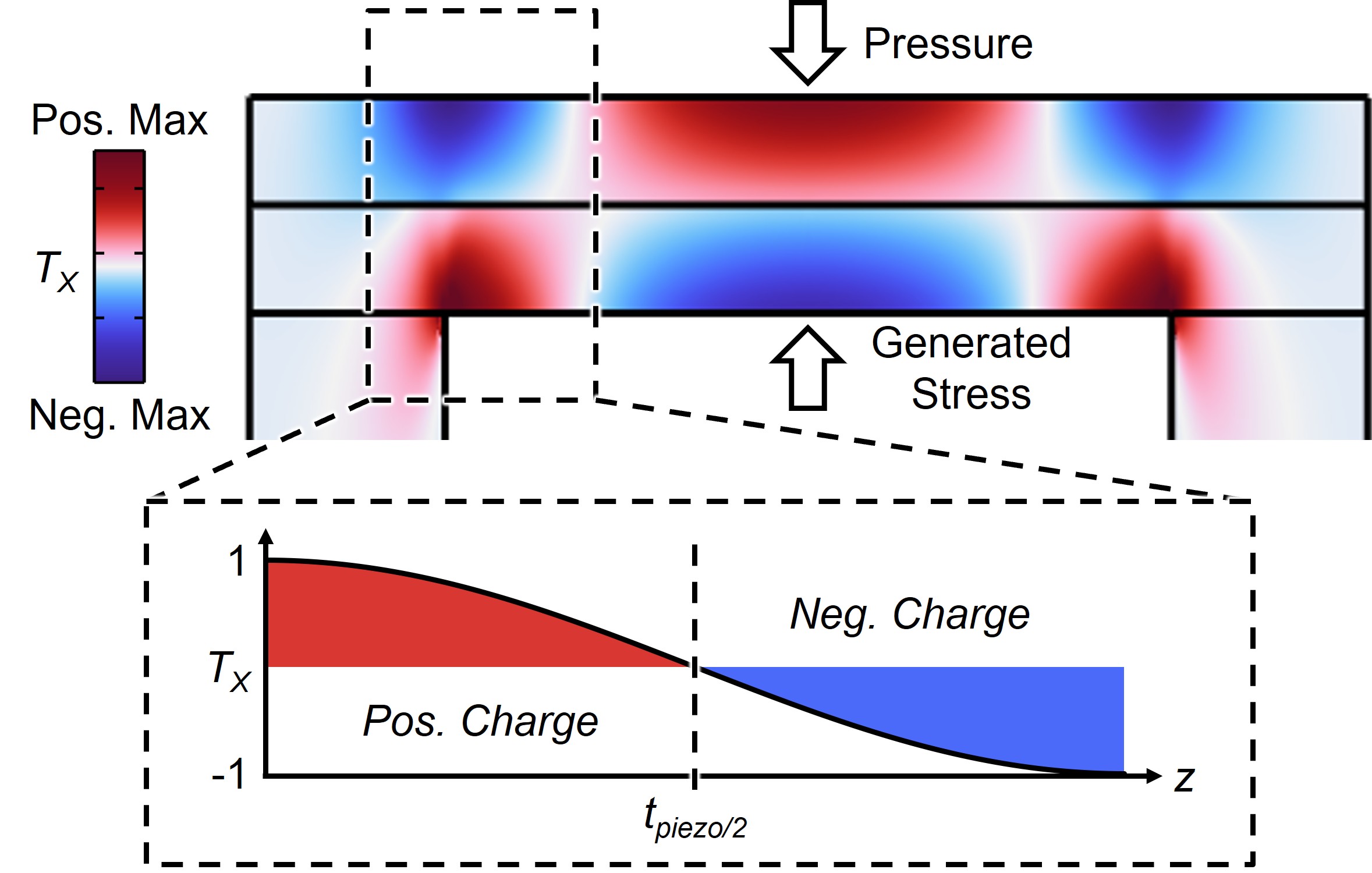}
    \caption{A cross-sectional view of the dominant in-plane stress profile characteristic of the flexural mode, highlighting the vertical charge cancellation due to the contrasting stress profile.}
    \label{fig:verticalcharge}
\end{figure}

\begin{figure}[!t]
    \centering
    \includegraphics[width=\columnwidth]{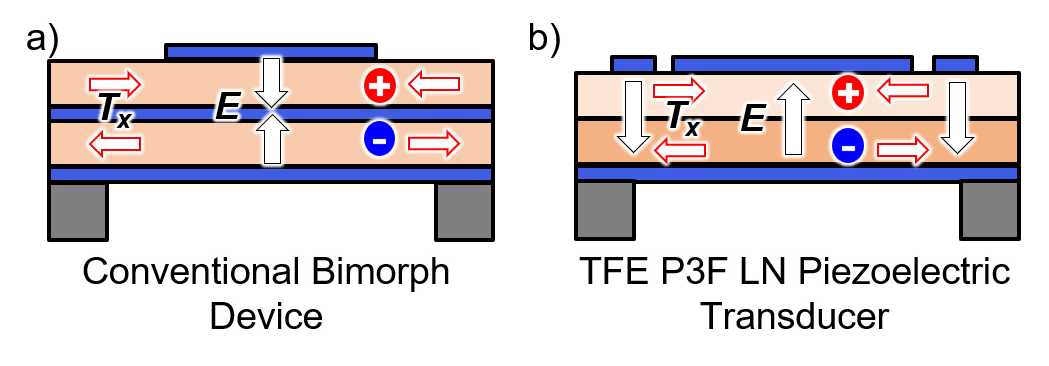}
    \caption{Cross-sectional views of (a) a conventional bimorph transducer and (b) the proposed TFE device using P3F LN.}
    \label{fig:bimorph_vs_tfe}
\end{figure}

State-of-the-art (SoA) PMUTs operate in the fundamental flexural mode, characterized by a contrasting dominant in-plane stress profile through the piezoelectric film thickness, requiring a multi-material stack for efficient operation. Fig.~\ref{fig:verticalcharge} shows how the in-plane stress evolves from tensile to compressive along the \(z\)-axis of the piezoelectric solid during flexural deflection, leading to internal charge cancellation in a single-layer (unimorph) piezoelectric solid.  To compensate, a passive layer can be utilized to displace the neutral stress axis in an effort to reduce destructive interference. However, a portion of the material's available performance is sacrificed as a result. As an alternative, a bilayer piezoelectric (bimorph) structure can be utilized, which has conventionally relied on an intermediate electrode layer in thickness-field-excited (TFE) devices to flip the electric field polarization between two layers of piezoelectric material, as shown in Fig.~\ref{fig:bimorph_vs_tfe}(a)~\cite{akhbari_bimorph_2016}. These structures harness the full potential of a given platform, at the cost of fabrication complexity and enhanced thermal and mechanical stress due to the presence of a large-area multi-material interface. 

Intrinsic limitations to bidirectional transduction and difficulties in high-quality bimorph stack fabrication have motivated research into alternative PMUT material platforms. Single-crystal lithium niobate (LN) has shown significant potential via its strong piezoelectric response and bimorph film availability~\cite{chulukhadze_bimorph_2026, zhao_piezoelectric_2025, yao_bimorph_2025}. More specifically, recent advances in LN thin-film transfer technologies have enabled bimorph periodically poled piezoelectric film (P3F) LN with single-crystal quality at film thicknesses ranging from as little as 100~nm to bulk values~\cite{chulukhadze_thickness-field_2026, kramer_thin-film_2023, kramer_acoustic_2025, lu_enabling_2020}. Here, constructive interference between two LN layers is achieved by flipping the relevant electromechanical coupling tensor components with either in-plane or out-of-plane rotation~\cite{kramer_generalized_2025}. Fig.~\ref{fig:bimorph_vs_tfe}(b) shows an example cross-section of a TFE P3F bimorph, optimized for robustness by removing the need for intermediate electrode layers. These developments have led to the demonstration of transferred P3F LN-based lateral-field-excited (LFE) transducers as a promising alternative to SoA PMUT platforms, showcasing high performance and resilience to temperatures as high as 800~\(^\circ\)C~\cite{chulukhadze_bimorph_2026, zhao_piezoelectric_2025, liu_high-temperature_2026}. However, the LFE mechanism showed a low static capacitance (\(C_0\)) density and employed a 20-\(\upmu\)m-thick active layer for robustness, limiting compliance and footprint for airborne applications. A TFE P3F prototype in~\cite{chulukhadze_thickness-field_2026} showed how the change in the electrical excitation scheme can overcome these limitations while preserving high performance, but its thickness severely limited the compliance.

This work aims to realize the potential shown by the initial bulk TFE P3F LN transducer prototype, complementing the LFE P3F LN platform by providing the first demonstration of a high-performance TFE P3F LN PMUT with a submicrometer-thick active layer, featuring a compact footprint and high performance fit for airborne applications.

\section{Design and Simulation}
\label{sec:design}

Devices were designed with a submicrometer-thick (670~nm) active layer to demonstrate TFE P3F transducers with high compliance and a compact footprint, optimal for sub-100-kHz airborne applications. The process began with LN orientation selection for TFE flexural-mode operation. Among commercially available LN material cuts with high relevant electromechanical coupling tensor components (\(e_{31,\mathrm{eff}}\) and \(e_{32,\mathrm{eff}}\)), 36Y and 128Y LN stand out as the primary candidates, as shown in Table~\ref{tab:thinfilm_ln}. Here, \(\varepsilon_{ii,\mathrm{eff}}\) and the ratio between \(e_{ij,\mathrm{eff}}\) and \(\varepsilon_{ii,\mathrm{eff}}\) are used as secondary performance metrics, where  \(\varepsilon_{ii,\mathrm{eff}}\) describes the \(C_0\)-density and the ratio describes the platform's open-circuit sensitivity, acting as another receiving performance metric. The comparison in Table~\ref{tab:thinfilm_ln} shows that while 36Y LN and 128Y LN exhibit electromechanical coupling tensor components of similar magnitude, the lower  \(\varepsilon_{ii,\mathrm{eff}}\) of 128Y LN yields a higher receiving FoM, leading to the adoption of 128Y LN for the TFE P3F LN transducer.

\begin{table}[!t]
\caption{Thin-film LN material parameters.}
\label{tab:thinfilm_ln}
\centering
\renewcommand{\arraystretch}{1.2}
\setlength{\tabcolsep}{4pt}
\small
\begin{tabular*}{\columnwidth}{@{\extracolsep{\fill}}ccccc}
\hline\hline
Material & \(i, j\) & \begin{tabular}{@{}c@{}}$e_{ij,\mathrm{eff}}$\\(C/m$^2$)\end{tabular} & $\varepsilon_{r,ii,\mathrm{eff}}$ & \begin{tabular}{@{}c@{}}\(e_{ij,\mathrm{eff}}/\varepsilon_{ii,\mathrm{eff}}\)\\(GV/m)\end{tabular} \\
\hline
Y-cut LN & 3, 1 & 3.1 & 46.9 & 7.5\\
36Y LN & 3, 1 & 3.5 & 50.7 & 7.8\\
128Y LN & 3, 1 & 3.6 & 37.6 & 10.8\\
\hline\hline
\end{tabular*}
\par\smallskip
\parbox{\columnwidth}{\footnotesize Magnitudes are listed; the sign of each coefficient depends on the orientation.}
\end{table}

\begin{figure}[!t]
    \centering
    \includegraphics[width=\columnwidth]{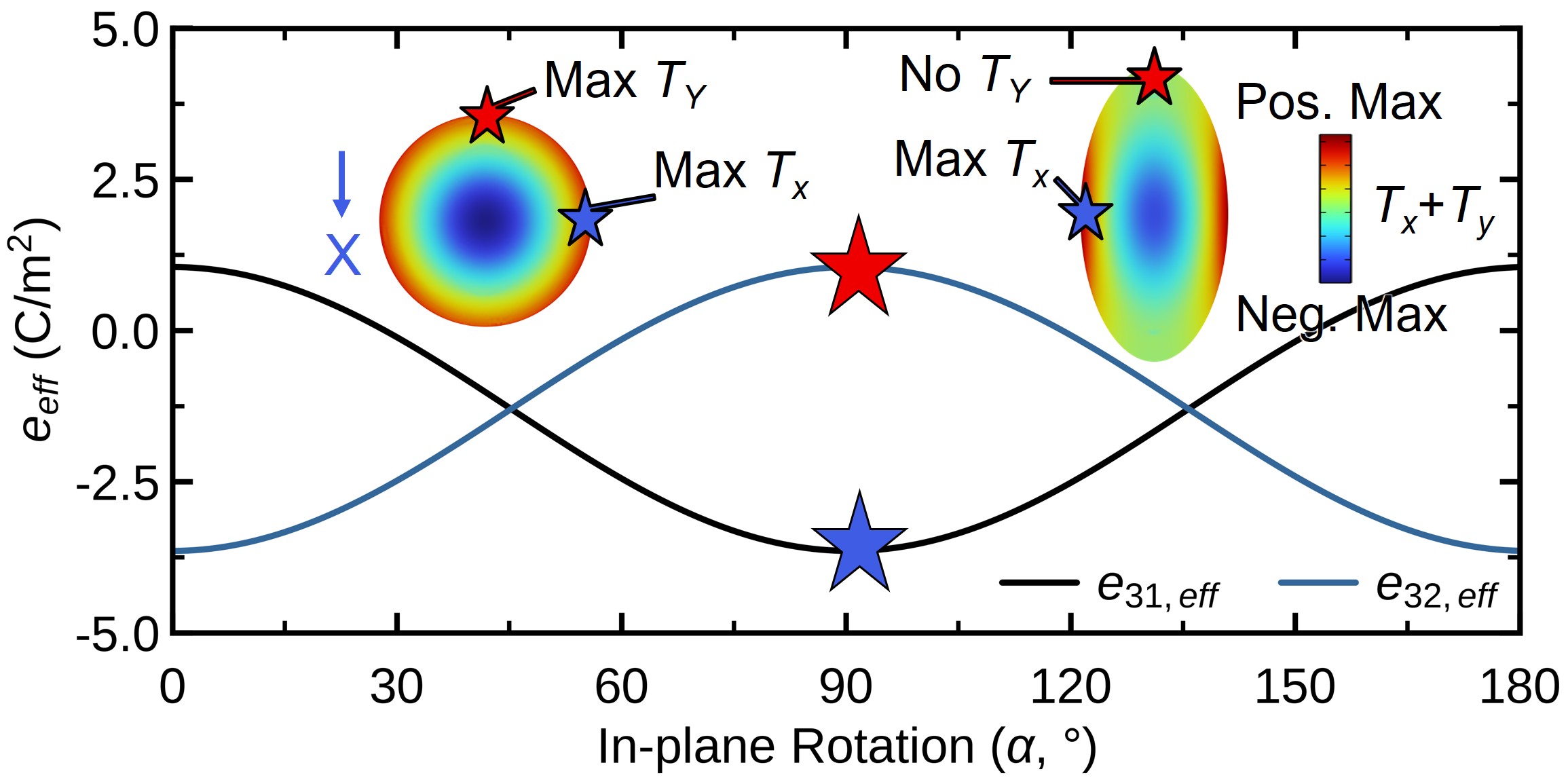}
    \caption{Relevant effective \(e\) electromechanical coupling tensor components plotted against rotation from the crystal X axis (\(\alpha\)), alongside top views of in-plane stress in a circular and an elliptical geometry. The results highlight how a circular geometry simultaneously excites an \(e_{31,\mathrm{eff}}\) of \(-3.6\)~C/m\(^2\) and \(e_{32,\mathrm{eff}}\) of \(+1.05\)~C/m\(^2\), while the elliptical geometry primarily harnesses \(e_{31,\mathrm{eff}}\) alone, alleviating destructive interference.}
    \label{fig:tfe_128vsinplane}
\end{figure}

\begin{figure}[!t]
    \centering
    \includegraphics[width=\columnwidth]{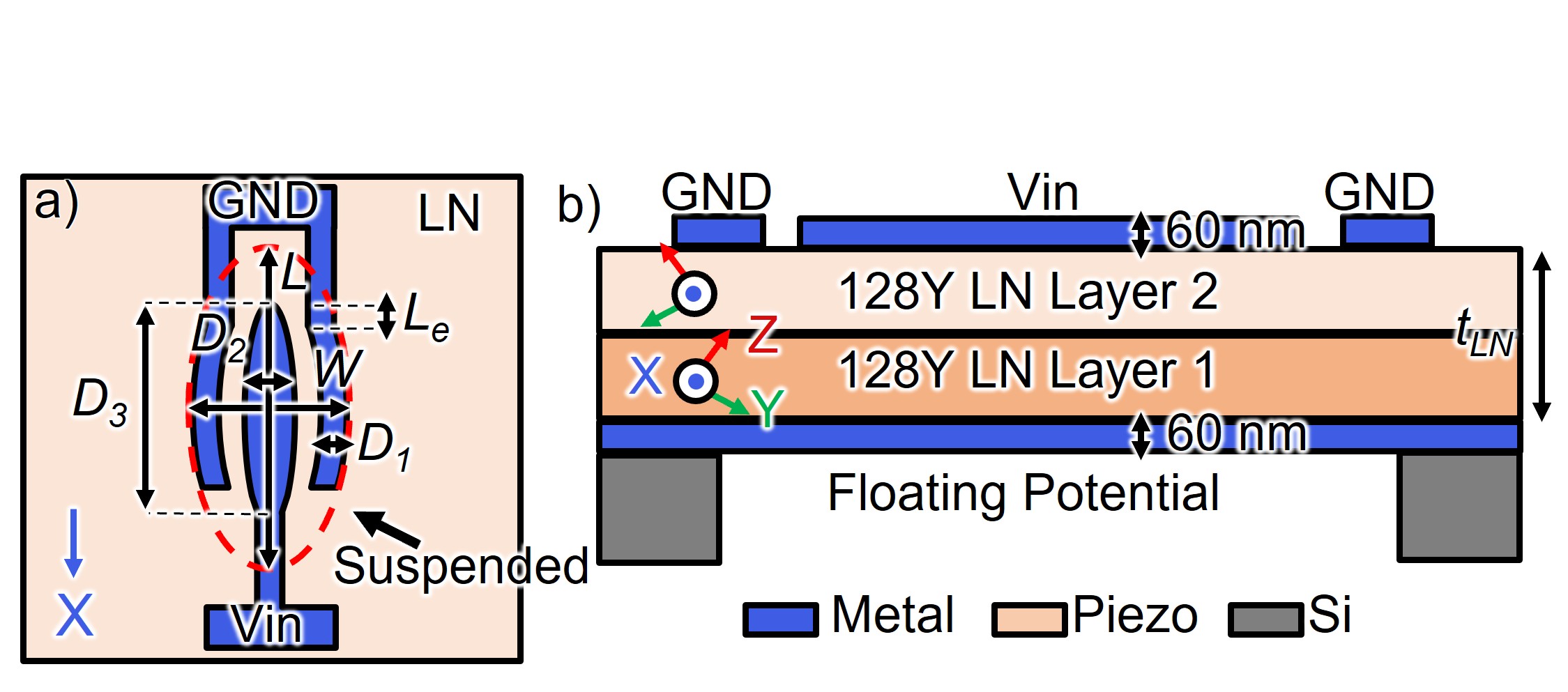}
    \caption{The (a) top and (b) cross-sectional views of the thin-film TFE P3F LN transducer, highlighting the elliptical membrane shape.}
    \label{fig:tfe_thinfilmtopviewcrosssection}
\end{figure}

Prior investigations into LN-based transducers showed the importance of in-plane anisotropy for flexural-mode operation~\cite{chulukhadze_bimorph_2026}. Hence, in an effort to study electrode placement, we plot the interaction between the effective in-plane electromechanical coupling tensor components (\(e_{31,\mathrm{eff}}\) and \(e_{32,\mathrm{eff}}\)), as shown in Fig.~\ref{fig:tfe_128vsinplane}. The analysis shows that the transverse \(e_{32,\mathrm{eff}}\) coupling coefficient is parasitic when \(e_{31,\mathrm{eff}}\) is at its peak value. Meanwhile, the inset mode shapes depict the dominant in-plane stress components during a flexural deflection, revealing significant transverse components in an isotropic, circular geometry. Hence, the in-plane anisotropy in 128Y LN necessitates a customized membrane shape designed and oriented to localize in-plane stress alongside the maximum \(e_{31,\mathrm{eff}}\) while eliminating the transverse stress antinodes. Prior studies on LFE P3F LN transducers encountered a similar symmetry in X-cut LN. They effectively utilized an elliptical membrane with the minor axis aligned with maximum \(e_{11,\mathrm{eff}}\) to minimize destructive interference~\cite{chulukhadze_bimorph_2026}. The same approach can be used here by aligning the minor axis of the ellipse to 90\(^\circ\) from the crystal X axis. Consequently, the top and the cross-sectional views of the thin-film TFE P3F LN transducer are shown in Fig.~\ref{fig:tfe_thinfilmtopviewcrosssection}(a) and (b), respectively, with the crystal axes and the key design parameters labeled. The design process closely followed that of the LFE counterpart in~\cite{chulukhadze_bimorph_2026}, using parametric sweeps while tracking the effective electromechanical coupling efficiency (\(k_{\mathrm{eff}}^{2}\)) and \(C_0\) defined as: 
\begin{subequations}
\label{eq:k2eff_C0}
\begin{align}
k_{\mathrm{eff}}^{2} &= \frac{\Phi^2 C_{\mathrm{ad}}}{C_0}, \label{eq:k2eff}\\
C_0 &= \frac{\operatorname{Im}(Y)}{\omega}, \label{eq:C0}
\end{align}
\end{subequations}
where \(\Phi\) is the turns ratio of the pressure-charge transformer, \(C_{\mathrm{ad}}\) is the acoustic compliance of the membrane, \(Y\) is the electrical admittance, and \(\omega\) is the angular frequency. Both \(\Phi\) and \(C_{\mathrm{ad}}\) are extracted under a unit pressure boundary load with the electrodes short-circuited: \(\Phi\) can be obtained as the short-circuit charge normalized by the volumetric displacement, and \(C_{\mathrm{ad}}\) is obtained by integrating the out-of-plane displacement over the membrane surface due to a unit pressure boundary load.

\begin{table}[!t]
\caption{Chosen key design parameters for the thin-film thickness-field-excited LN transducer.}
\label{tab:tfe_geom_params}
\centering
\renewcommand{\arraystretch}{1.2}
\setlength{\tabcolsep}{2pt}
\small
\begin{tabular*}{\columnwidth}{@{\extracolsep{\fill}}ccccccc}
\hline
$t_{\mathrm{LN}}$& $W$ & $L$ & $D_1$ & $D_2$ & $D_3$ & $L_e$ \\
\hline
0.67~\(\upmu\)m & 400~\(\upmu\)m & 800~\(\upmu\)m & 37.5~\(\upmu\)m & 120~\(\upmu\)m & 690~\(\upmu\)m & 115~\(\upmu\)m \\
\hline
\end{tabular*}
\end{table}

The proposed structure relies on a floating bottom electrode while utilizing the top electrodes for signal and ground, thereby establishing thickness-field excitation under electrostatic equilibrium, alleviating fabrication-related challenges associated with a patterned bottom electrode~\cite{chulukhadze_toward_2026, chulukhadze_miniature_2025, yao2026bimorphlithiumniobatethicknessshear, chulukhadze_planar_2025}. This design choice limits the achievable \(C_0\) of the device by placing the inner and outer electrodes in series, which is an acceptable design trade-off due to the moderate \(\varepsilon_{r,ii,\mathrm{eff}}\) in LN. 

The obtained key variables, summarized in Table~\ref{tab:tfe_geom_params}, aim to maximize performance for a device with a resonant frequency following the audio band (20~Hz--20~kHz) at 30~kHz, indicating an exceedingly compact footprint of 0.25~mm\(^2\) for sub-100-kHz airborne operation.

\begin{figure}[!t]
    \centering
    \includegraphics[width=\columnwidth]{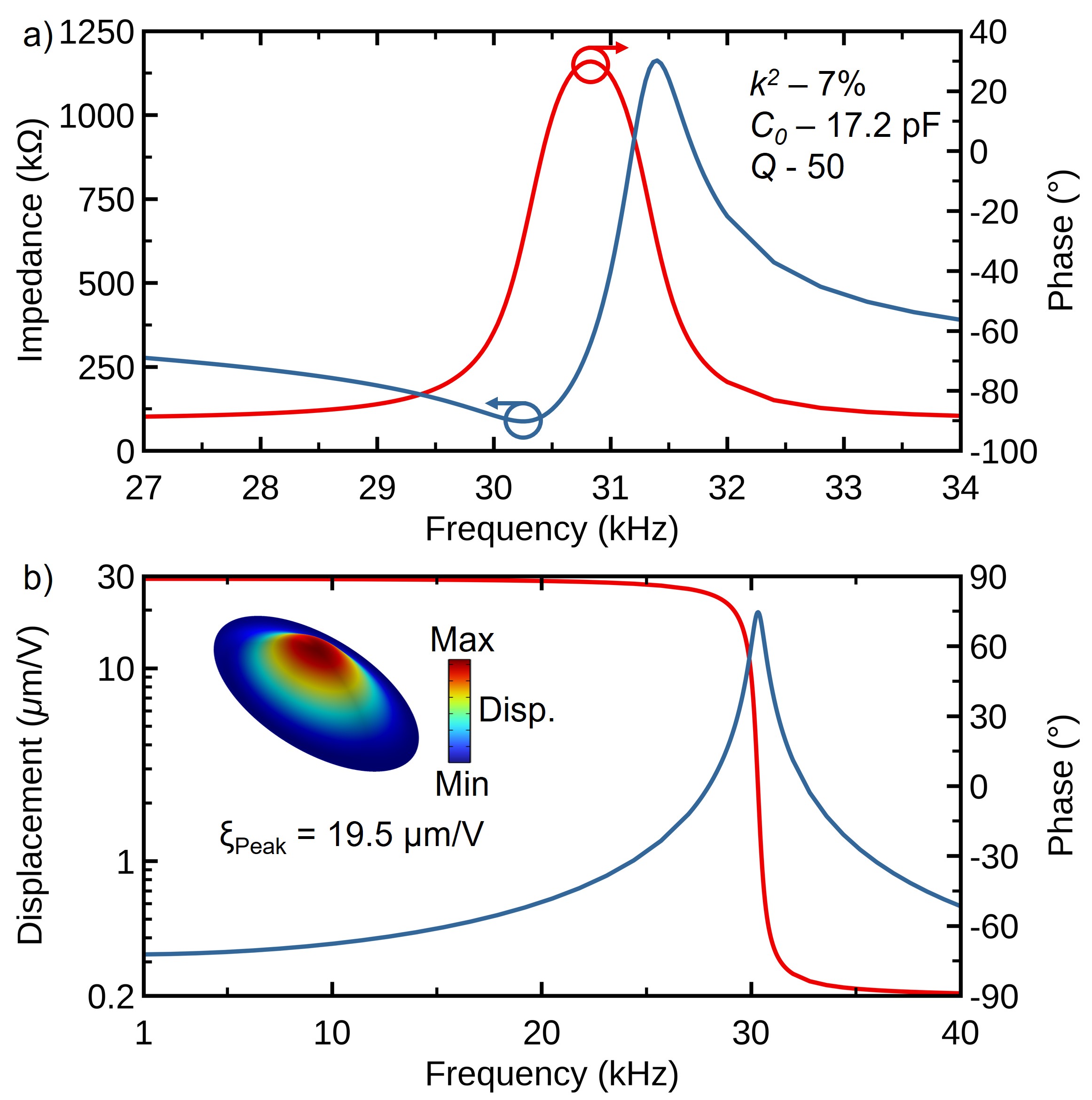}
    \caption{FEA-simulated (a) electrical impedance and phase, and (b) displacement sensitivity for a nominal quality factor of 50. Extracted performance parameters are labeled.}
    \label{fig:tfe_thinfilmsimulations}
\end{figure}

The device was simulated in vacuum using COMSOL finite-element analysis (FEA) to extract the electromechanical performance, decoupling significant acoustic effects that may arise with a compliant piezoelectric membrane. The resulting impedance and displacement plots can be seen in Fig.~\ref{fig:tfe_thinfilmsimulations}(a) and (b), with an assumed quality factor (\(Q\)) of 50. The device achieves a high \(k_{\mathrm{eff}}^{2}\) of 7\% alongside a \(C_0\) of 17.2~pF and a high peak out-of-plane displacement (\(\xi_{\mathrm{peak}}\)) of 19.5~\(\upmu\)m/V at 30.4~kHz. In order to decouple resonant effects for flat-band operation, the quasi-static peak out-of-plane displacement (\(\xi_{\mathrm{static}}\)) can be evaluated as close to DC as possible, yielding 330~nm/V at 1~kHz, indicating high film compliance in combination with a high \(k_{\mathrm{eff}}^{2}\). The simulated device demonstrates an excellent \(k_{\mathrm{eff}}^{2}\) alongside a high \(\xi_{\mathrm{static}}\) in a compact footprint of 0.25~mm$^2$ while achieving a moderate \(C_0\) of 17.2~pF, in spite of the floating bottom electrode. Hence, the thin-film TFE P3F LN platform features high performance on par with its LFE counterpart and shows good integration capability via its \(C_0\)-density. 

\section{Fabrication and Measurement}
\label{sec:fabmeas}

\begin{figure}[!t]
    \centering
    \includegraphics[width=\columnwidth]{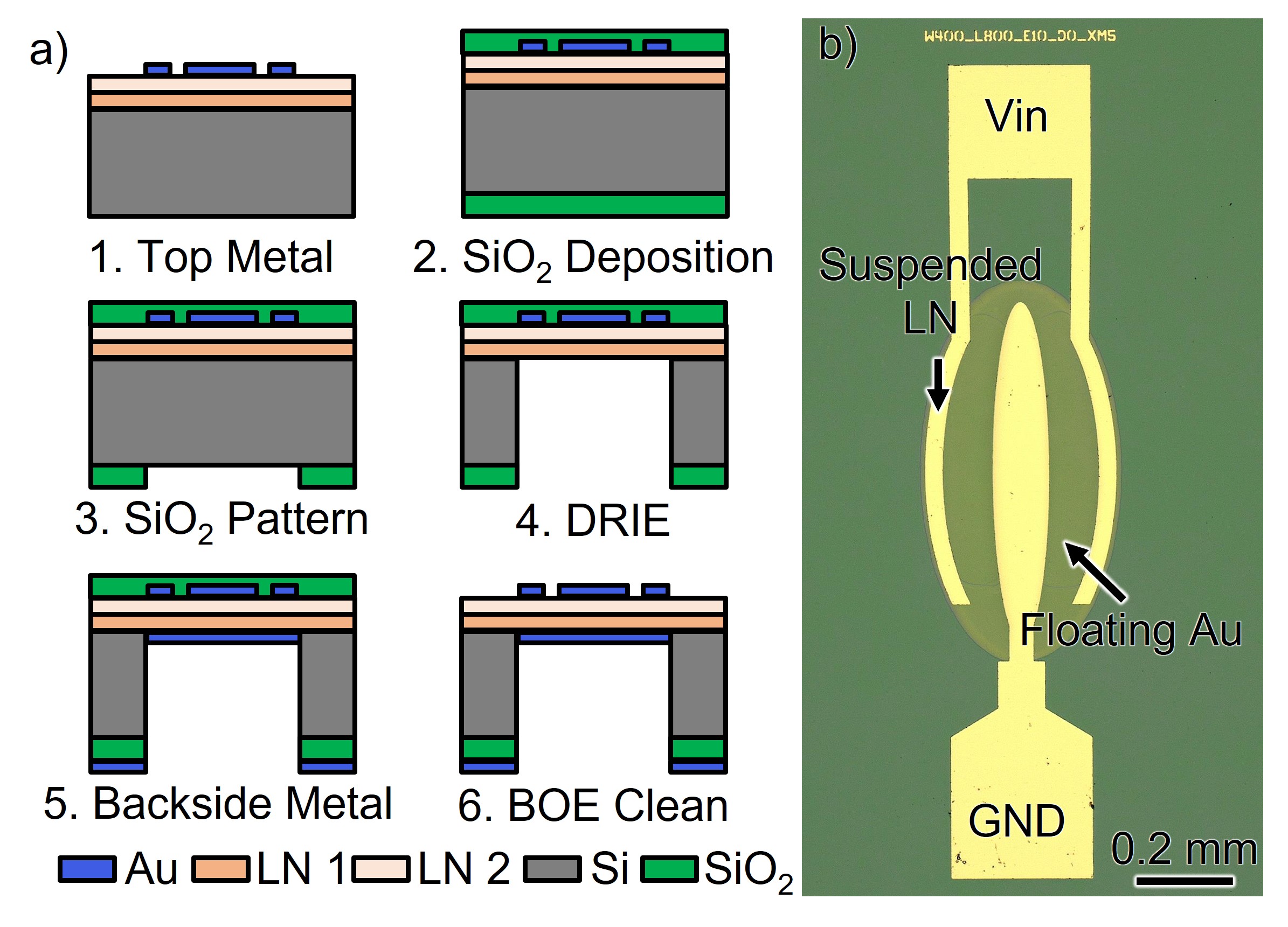}
    \caption{(a) The fabrication flow of the thin-film TFE transducer. (b) An optical image of the resulting device.}
    \label{fig:tfe_thinfilm_faboptical}
\end{figure}

The transducers were fabricated using a 670-nm-thick 128Y P3F LN wafer, provided by NGK Corporation. The fabrication began by evaporating 50~nm of gold (Au) top electrodes with a 10-nm-thick chromium (Cr) adhesion layer, introducing minimal mass loading while featuring high conductivity. Next, silicon dioxide (SiO$_2$) was deposited on the top and back sides of the wafer using low-temperature plasma-enhanced chemical vapor deposition (PECVD) at 100~\(^\circ\)C to act as a protective layer for the active material as well as the substrate during the deep reactive ion etching (DRIE) process. The thicknesses were 200~nm and 1~\(\upmu\)m for the top and the bottom layers, respectively. The etched trench and its alignment with the top electrodes were evaluated optically due to LN's transparency. Once a satisfactory result was reached, 50~nm Au alongside a 10-nm Cr adhesion layer was deposited via electron-beam evaporation into etched Si trenches to form the bottom electrode. A buffered oxide etch (BOE) was used to clean the remaining SiO$_2$ residue. The fabrication flow and an optical image of the device can be seen in Fig.~\ref{fig:tfe_thinfilm_faboptical}(a) and (b).

\begin{figure}[!t]
    \centering
    \includegraphics[width=\columnwidth]{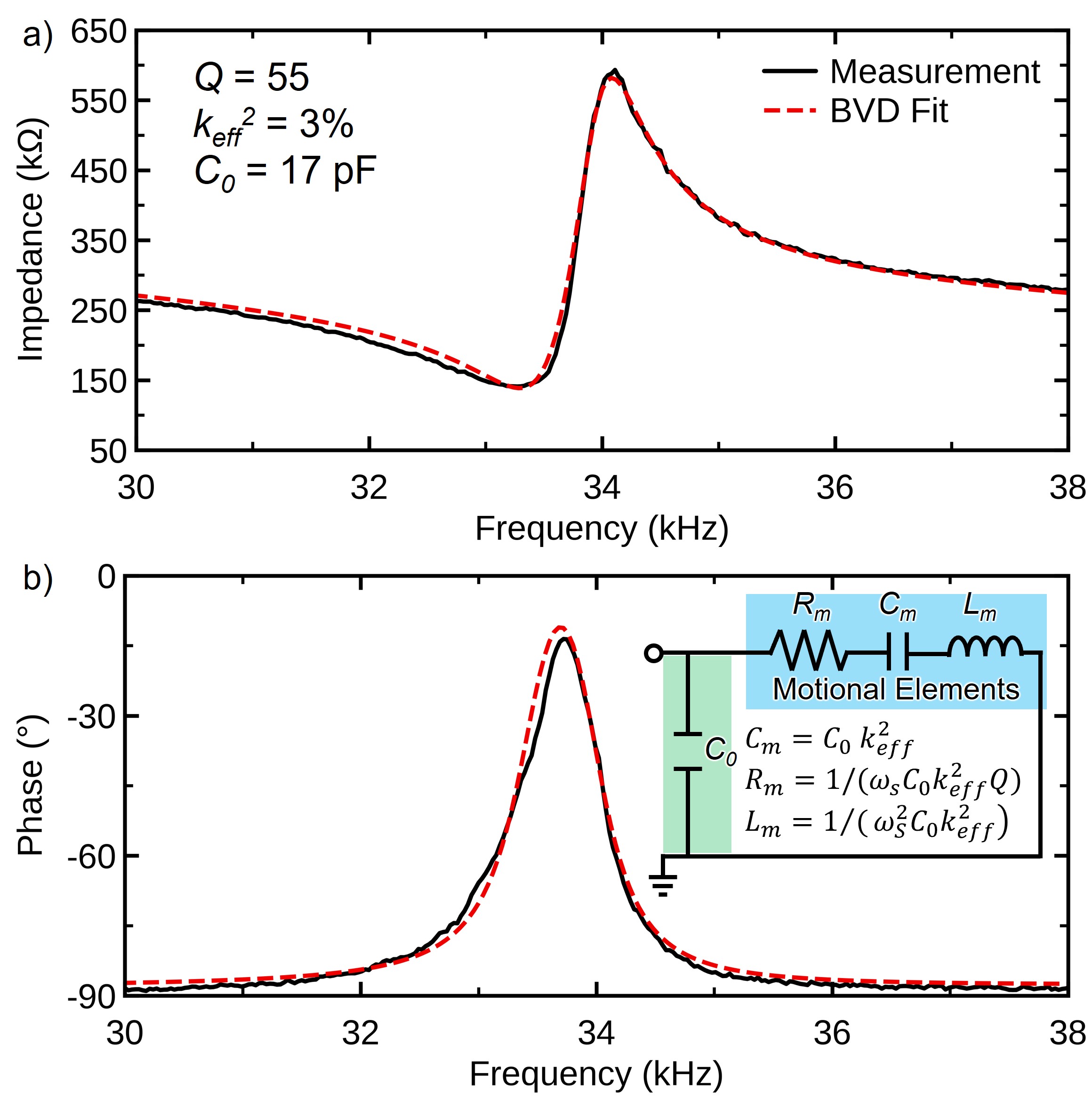}
    \caption{Measured (a) impedance and (b) phase of the TFE P3F LN transducer on a grounded elevated plate, alongside its BVD equivalent circuit model fit (inset).}
    \label{fig:tfe_elevatedelecmeas}
\end{figure}

\begin{figure}[!t]
    \centering
    \includegraphics[width=\columnwidth]{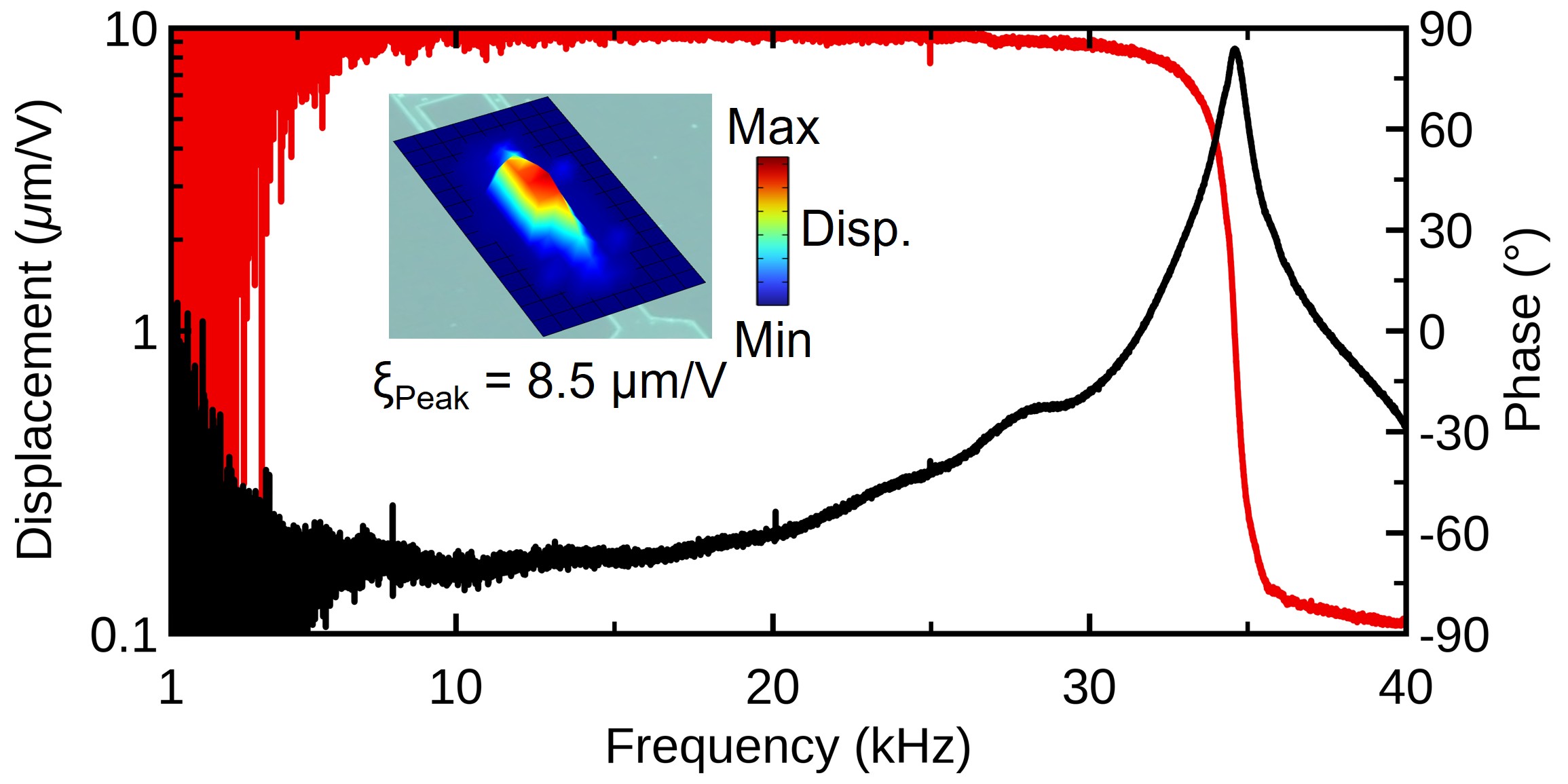}
    \caption{The LDV-measured out-of-plane displacement and phase of the transducer on an elevated, grounded metallic plate, showcasing a high transmit efficiency with a \(\xi_{\mathrm{peak}}\) of 8.5~\(\upmu\)m/V and \(\xi_{\mathrm{static}}\) of 145~nm/V.}
    \label{fig:tfe_elevatedmechmeas}
\end{figure}

The fabricated device was measured with a Zurich MFIA impedance analyzer and a Polytec laser Doppler vibrometer (LDV), with the results shown in Figs.~\ref{fig:tfe_elevatedelecmeas} and \ref{fig:tfe_elevatedmechmeas}, respectively. The measurements were taken in air on an elevated, grounded metallic pedestal with a cutout section under the device in an effort to minimize the influence of the back cavity on the transducer. An equivalent Butterworth--Van Dyke (BVD) circuit model was utilized to extract device performance, shown as an inset image in Fig.~\ref{fig:tfe_elevatedelecmeas}(b), where the motional components are parametrized such that the extracted \(k_{\mathrm{eff}}^{2}\) follows a definition identical to \eqref{eq:k2eff}. The resulting data revealed a well-defined resonance at 33.4~kHz with an extracted \(k_{\mathrm{eff}}^{2}\) of 3\% and a \(C_0\) of 17~pF, in line with the simulated value from Fig.~\ref{fig:tfe_thinfilmsimulations}. The LDV measurement in Fig.~\ref{fig:tfe_elevatedmechmeas} excites the device with a 50-mV periodic chirp, confirming fundamental flexural-mode operation and yielding a high \(\xi_{\mathrm{peak}}\) of 8.5~\(\upmu\)m/V at resonance. The audio band can be analyzed for \(\xi_{\mathrm{static}}\); however, the response near DC contains significant artifacts from the chirped signal, and \(\xi_{\mathrm{static}}\) is therefore obtained at 5~kHz, which is still in the flat band. The measurement yields a high \(\xi_{\mathrm{static}}\) of 145~nm/V, which will be used to evaluate device performance within the audio band. 

In tandem, the measured results indicate that the device obtains an excellent transmission efficiency and exhibits a low resonance frequency ideal for ultrasonic and audio-band applications while featuring an exceedingly compact footprint of 0.25~mm\(^2\). Moreover, the suspended film consists of only a 670-nm-thick active layer, demonstrating the ability of transferred thin-film P3F LN to realize large-area membranes. Regardless, the presented data warrants further discussion, as important metrics such as receiving sensitivity, acoustic effects, and simulation discrepancies remain to be investigated.

\section{Discussion}
\label{sec:discussion}

\subsection{Sensitivity}
\label{subsec:sensitivity}

Using the LDV measurement in Fig.~\ref{fig:tfe_elevatedmechmeas} and the BVD model parameters from Fig.~\ref{fig:tfe_elevatedelecmeas}, both the charge and voltage sensitivity of the transducer can be extrapolated. Charge sensitivity (\(S_{\mathrm{SC}}\), in C/Pa) can be directly obtained through the measured two-dimensional out-of-plane displacement mode shape as follows:
\begin{equation}
    S_{\mathrm{SC}} = \iint_{A} w \,\mathrm{d}A,
    \label{eq:ssc}
\end{equation}
where $w$ is the out-of-plane particle displacement per volt, measured as close to DC as possible (here, 5~kHz) to stay within the flat band well below resonance, and $A$ is the membrane area. Hence, \eqref{eq:ssc} shows that the volumetric displacement per volt is equivalent to the stored charge per unit pressure, which stems from the inherent reciprocity within the piezoelectric constitutive equation system~\cite{hunt_electroacoustics_1982, prasad_analytical_2006}. Here, analysis at 5~kHz yields a moderate \(S_{\mathrm{SC}}\) of 24.7~fC/Pa, which can be used as an effective comparison metric against other airborne transducers. The open-circuit sensitivity (\(S_{\mathrm{OC}}\), in V/Pa) can be obtained by dividing \(S_{\mathrm{SC}}\) by the free capacitance of the system, found as $C_{\mathrm{m}} + C_0$ from Fig.~\ref{fig:tfe_elevatedelecmeas}(b)~\cite{prasad_analytical_2006}. Accordingly, \(S_{\mathrm{OC}}\) at 5~kHz is 1.4~mV/Pa, indicating a high voltage sensitivity within the flat band. Regardless, these metrics are strongly influenced by both acoustic and system-level effects, such as the back-cavity compliance (\(C_{\mathrm{b}}\)) and input capacitance of an amplifier~\cite{seo_theoretical_2017}. While the system-level integration of the device is out of the scope of this work, the influence of the back cavity on both actuating and sensing performance will be discussed next. 

\subsection{Cavity Effects}
\label{subsec:cavityeffects}

\begin{figure}[!t]
    \centering
    \includegraphics[width=\columnwidth]{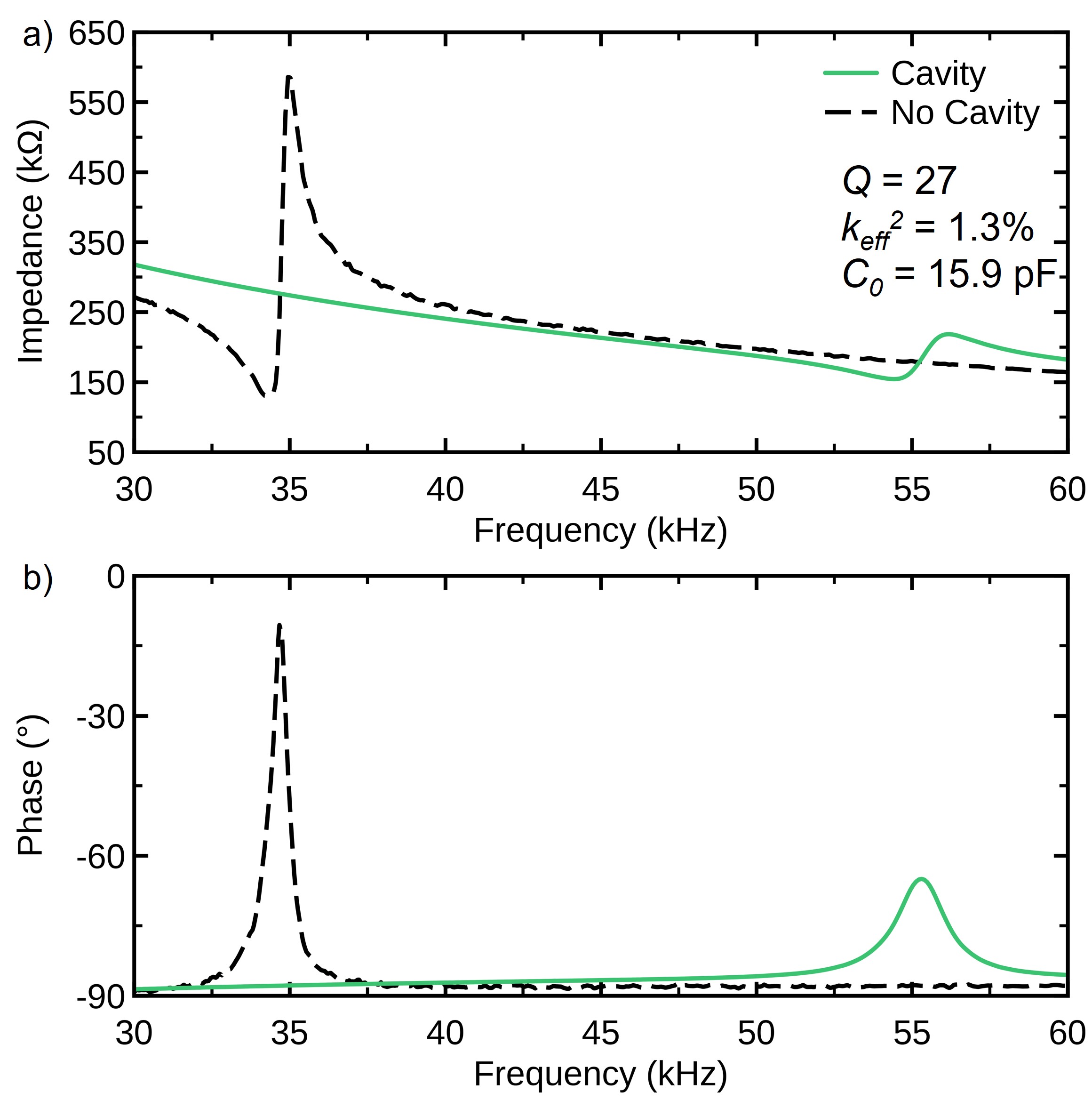}
    \caption{The impedance measurement of the device with the back-cavity effect included, compared with the measurements taken on an elevated pedestal, highlighting the frequency offset as well as changing damping and electromechanical performance.}
    \label{fig:tfe_cavityvselevatedelec}
\end{figure}

\begin{figure}[!t]
    \centering
    \includegraphics[width=\columnwidth]{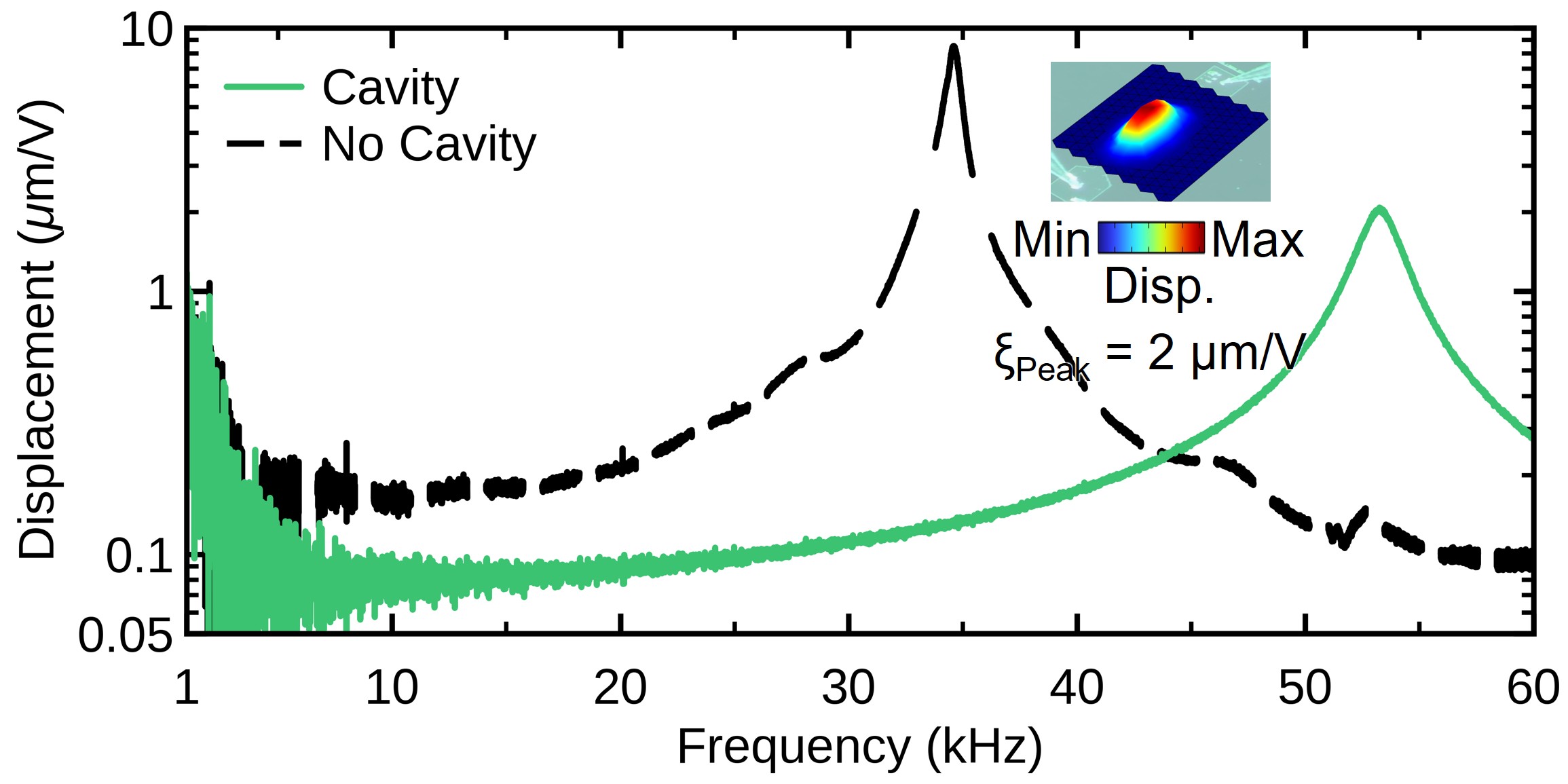}
    \caption{LDV-measured out-of-plane displacement response of the transducer with the back-cavity effect present, compared with the original measurement, indicating a reduced \(\xi_{\mathrm{peak}}\) as well as a reduced \(\xi_{\mathrm{static}}\).}
    \label{fig:tfe_cavityvselevatedmech}
\end{figure}

In an effort to investigate back-cavity-induced acoustic effects, the elevated pedestal was exchanged for a flat metallic substrate. The transducer showed a significant change in performance and resonant behavior, with the comparison shown in Figs.~\ref{fig:tfe_cavityvselevatedelec} and \ref{fig:tfe_cavityvselevatedmech}. The measured impedance indicates a frequency upshift due to \(C_{\mathrm{b}}\), while also showing a reduced \(k_{\mathrm{eff}}^{2}\), \(Q\), and lowered \(C_0\) when fit with the BVD model from Fig.~\ref{fig:tfe_elevatedelecmeas}(b). On the other hand, the LDV measurement shows both a reduced \(\xi_{\mathrm{peak}}\) and a reduced \(\xi_{\mathrm{static}}\), indicating the lower overall compliance of the system. These results can be used to extrapolate the receiving performance with the influence of the back cavity present; however, an important distinction must be made: the BVD fit of the electrical measurement in Fig.~\ref{fig:tfe_cavityvselevatedelec} lumps \(C_{\mathrm{b}}\) into \(C_{\mathrm{m}}\), although \(C_{\mathrm{b}}\) does not contribute to the piezoelectric effect, and this approach can significantly underestimate \(C_{\mathrm{m}}\), and hence \(k_{\mathrm{eff}}^{2}\), when in series with \(C_{\mathrm{b}}\) of comparable value. The LDV measurement can be analyzed to extract sensitivity metrics using \eqref{eq:ssc}. Accordingly, \(S_{\mathrm{SC}}\) with the back-cavity effect present is found to be 12.5~fC/Pa at 5~kHz, while \(S_{\mathrm{OC}}\) is 0.78~mV/Pa at the same frequency, presenting a twofold reduction from the original measurement. Hence, the back cavity plays a critical role in determining the transducer's performance when the membrane compliance is sufficiently high. Future design iterations must aim to either minimize its effect by enhancing \(C_{\mathrm{b}}\) compared to the membrane compliance, or closely match them~\cite{seo_theoretical_2017, horowitz_development_2007}.

\subsection{Comparison to Prior Works}
\label{subsec:literaturecomparison}


\begin{table}[!t]
\caption{Survey of SoA sub-100-kHz air-coupled piezoelectric transducers, highlighting the performance and the compact footprint of the presented device.}
\label{tab:sota_survey}
\centering
\renewcommand{\arraystretch}{1.2}
\setlength{\tabcolsep}{1.5pt}
\small
\begin{tabular*}{\columnwidth}{@{\extracolsep{\fill}}cccccccc}
\hline
Material & $f_{\mathrm{s}}$ & Area & Thickness$^{\#}$ & $\xi_{\mathrm{static}}$ & $C_{0}$ & $S_{\mathrm{SC}}^{*}$ & $S_{\mathrm{OC}}^{*}$ \\
 & (kHz) & (mm$^{2}$) & (\(\upmu\)m) & (nm/V) & (pF) & (fC/Pa) & (mV/Pa) \\
\hline
PZT \cite{luo_airborne_2021} & 40 & 1.29 & 8.4 & 430$^{\dagger}$ & 1200 & 96 & 0.08$^{\dagger}$ \\
ScAlN \cite{trivedi_beam_2025} & 47.9 & 2.54 & 10.6 & 33 & 256 & 82$^{\ddagger}$ & 0.32$^{\dagger}$ \\
ScAlN \cite{hu_sc0096_2024} & 30.7 & 0.53 & 1.3 & -- & 107 & 38$^{\ddagger}$ & 0.35 \\
\tw{This Work} & \tw{33.4} & \tw{0.25} & \tw{0.79} & \tw{145} & \tw{17} & \tw{25} & \tw{1.4} \\
\hline
\end{tabular*}
\par\smallskip
\parbox{\columnwidth}{\footnotesize $^{\#}$Includes the electrode thickness. $^{*}$In the flat band. $^{\dagger}$Resonant value divided by the reported $Q$. $^{\ddagger}$$S_{\mathrm{SC}} = S_{\mathrm{OC}}C$ with the measured capacitance.}
\end{table}

The extracted device performance is compared to SoA airborne piezoelectric transducers with a resonance frequency below 100~kHz in Table~\ref{tab:sota_survey}~\cite{luo_airborne_2021, trivedi_beam_2025, hu_sc0096_2024}. The comparison focuses on evaluating the balance between \(\xi_{\mathrm{static}}\) and audio-band sensitivity metrics. The comparison highlights that the presented LN device is the most compact of the surveyed flexural-mode clamped membrane transducers with a sub-100-kHz resonance frequency, while also featuring the highest \(S_{\mathrm{OC}}\) alongside a moderate \(\xi_{\mathrm{static}}\) comparable to high-transmit-efficiency material platforms, such as PZT. This performance is achieved with the combination of the submicrometer-thick bimorph active layer and the achieved \(k_{\mathrm{eff}}^{2}\), facilitating a high volumetric displacement in the flat band, underscoring the potential of the TFE P3F LN transducer platform.

The single-crystal PZT transducer in~\cite{luo_airborne_2021} obtains a high flat-band \(S_{\mathrm{SC}}\) of 96~fC/Pa owing to a \(C_0\) of 1200~pF, which also limits its \(S_{\mathrm{OC}}\) to 0.08~mV/Pa, while featuring a large \(\xi_{\mathrm{static}}\) of 430~nm/V. On the other hand, a clamped 30\% ScAlN membrane from~\cite{trivedi_beam_2025} simultaneously obtains a high flat-band \(S_{\mathrm{SC}}\) of 82~fC/Pa and a high \(S_{\mathrm{OC}}\) of 0.32~mV/Pa, but shows a limitation in its \(\xi_{\mathrm{static}}\). Its 9.6\% ScAlN bimorph counterpart in~\cite{hu_sc0096_2024} obtains similar performance despite its lower scandium alloy incorporation, yielding an \(S_{\mathrm{SC}}\) of 38~fC/Pa and \(S_{\mathrm{OC}}\) of 0.35~mV/Pa. In comparison with these devices, the TFE P3F LN PMUT strikes a balance between transmitting and receiving performance: the measured \(\xi_{\mathrm{static}}\) is comparable with PZT and significantly higher than ScAlN. Its \(S_{\mathrm{SC}}\) is competitive with both despite the smaller \(C_0\) of the LN transducer, while its flat-band \(S_{\mathrm{OC}}\) is the highest. Importantly, the \(S_{\mathrm{SC}}\) can be further enhanced by exchanging the floating bottom electrode architecture for its grounded version, maximizing \(C_0\) density.

The presented TFE P3F LN results highlight the platform's ability to realize large-area membranes with a high \(k_{\mathrm{eff}}^{2}\), enabling a high \(S_{\mathrm{OC}}\) alongside a high \(\xi_{\mathrm{static}}\). Compared with prior demonstrations of P3F LN transducers, summarized in Table~\ref{tab:bimorph_survey}, this implementation introduces an additional application space. The LFE P3F LN PMUT in~\cite{zhao_piezoelectric_2025} shows a balance between robustness and performance, utilizing a 2-\(\upmu\)m bimorph for an LFE P3F transducer with a small footprint fit for array applications in the MHz regime. A similar LFE P3F LN transducer in~\cite{chulukhadze_bimorph_2026} scaled the film thickness to 20~\(\upmu\)m, featuring robustness to extreme environmental conditions. The TFE P3F LN prototype in~\cite{chulukhadze_thickness-field_2026} showed promise for airborne applications, which was realized in this work using a 670-nm-thick P3F active layer. With this addition, the P3F LN platform demonstrates exceptional flexibility with its ability to realize high-quality films for piezoelectric transducers with thicknesses ranging from submicrometer to bulk values.

\begin{table}[!t]
\caption{Survey of prior P3F LN-based flexural-mode piezoelectric transducers, highlighting the submicrometer film thickness of the presented implementation, and underscoring the platform's viability across frequencies and applications.}
\label{tab:bimorph_survey}
\centering
\renewcommand{\arraystretch}{1.2}
\setlength{\tabcolsep}{2pt}
\footnotesize
\begin{tabular*}{\columnwidth}{@{\extracolsep{\fill}}lccccccc}
\hline
Work & Material & Interlayer & Exci- & $t_{\mathrm{act}}$ & $f_{\mathrm{s}}$ & Area & $k_{\mathrm{eff}}^{2}$ \\
 & & electrode & tation & (\(\upmu\)m) & (kHz) & (mm$^{2}$) & (\%) \\
\hline
\cite{zhao_piezoelectric_2025} & X LN & N & LFE & 2 & 1200 & 0.04 & 3.7 \\
\cite{chulukhadze_bimorph_2026} & X LN & N & LFE & 20 & 775 & 0.28 & 6.4 \\
\cite{chulukhadze_thickness-field_2026} & 36Y LN & N & TFE & 300 & 115 & 23.0 & 5.0 \\
\tw{This Work} & \tw{128Y LN} & \tw{N} & \tw{TFE} & \tw{0.67} & \tw{33.4} & \tw{0.25} & \tw{3.0} \\
\hline
\end{tabular*}
\end{table}


\begin{figure}[!t]
    \centering
    \includegraphics[width=\columnwidth]{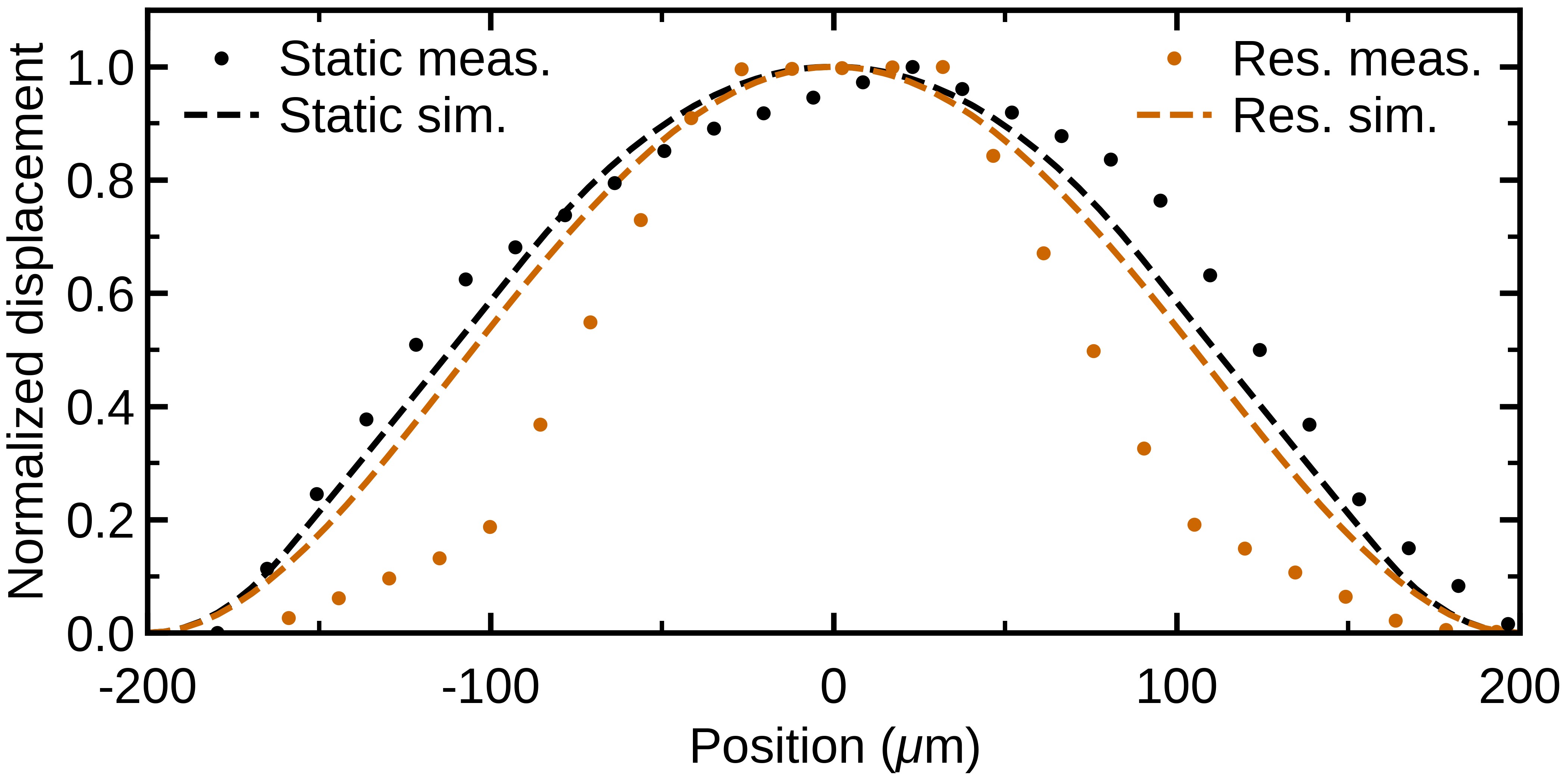}
    \caption{Comparison of simulated and measured flexural mode shapes on a cut line defined on the minor axis of the ellipse in the flat band and at resonance, highlighting that the fabricated device does not follow the conventional mode shape of a clamped plate. Static data reveal a wider displacement distribution than expected, while the resonance data indicate the opposite effect, indicating a major shift in boundary conditions.}
    \label{fig:tfe_modeshapeanalysis}
\end{figure}

While the fabricated device exhibits high performance compared with incumbent airborne PMUTs, the analysis of the obtained measurements against the idealized simulation in Fig.~\ref{fig:tfe_thinfilmsimulations} shows that the device has not yet realized its full potential.  The comparison between Figs.~\ref{fig:tfe_thinfilmsimulations} and \ref{fig:tfe_elevatedelecmeas} shows that the measured device obtained a lower \(k_{\mathrm{eff}}^{2}\) than predicted, leading to a significantly reduced \(\xi_{\mathrm{peak}}\) and \(\xi_{\mathrm{static}}\). At the same time, the device exhibited a resonance frequency approximately 3~kHz higher than in the simulation. The LDV-measured mode shape offers a direct explanation when compared with its simulated version, shown in Fig.~\ref{fig:tfe_modeshapeanalysis}. The contrast between the ideal simulated mode shape of a clamped membrane and the measured data indicates that the fabricated LN device experiences a fundamental shift in boundary conditions, demonstrating a flexural mode shape with a significantly wider displacement distribution in the flat band, and a narrower shape at resonance. This discrepancy between simulation and measurement describes a complex effect that has a profound influence on device design and performance and will be the primary focus of future studies on thin-film P3F LN-based transducers.

\section{Conclusion}
This work presents the first demonstration of a P3F LN-based submicrometer-thick, clamped-plate piezoelectric transducer operating at sub-100-kHz frequencies for airborne applications. The device achieves an exceedingly compact footprint due to the high film compliance achieved by the large-area, thin-film membrane, which consequently leads to high electromechanical performance in conjunction with the high \(e_{31,\mathrm{eff}}\) in single-crystal LN. As a result, the fabricated device demonstrates a \(k_{\mathrm{eff}}^{2}\) of 3\% with a \(C_0\) of 17~pF, a resonance frequency of 33.4~kHz, and yields a high \(\xi_{\mathrm{static}}\), as well as a moderate extracted flat-band \(S_{\mathrm{SC}}\) and a high flat-band \(S_{\mathrm{OC}}\). Compared with SoA airborne piezoelectric transducers, the LN device achieves a comparable \(\xi_{\mathrm{static}}\) while showcasing a significantly higher \(S_{\mathrm{OC}}\), positioning LN as a SoA sensor and bidirectional transducer platform. Importantly, the presented study underscores further potential of the TFE P3F LN transducer platform by pinpointing discrepancies between the simulation and the measured results, outlining clear next steps in order to realize the potential performance of the P3F LN piezoelectric transducer platform. 

\section*{Acknowledgment}
The authors would like to thank Dr.~Todd Bauer for helpful discussions.

\balance

\bibliographystyle{IEEEtran}
\bibliography{main}

\vfill
\end{document}